\documentclass{sig-alternate}
\usepackage{balance}

\usepackage{threeparttable}
\usepackage{caption}
\usepackage{xspace}
\usepackage{amsmath}
\usepackage{graphicx}
\usepackage{subcaption}
\usepackage{float}

\newcommand{\para}[1]{\noindent\textbf{#1}}
\newcommand{\preint}{predicted velocity range\xspace}

\begin{document}

% Copyright
\setcopyright{acmcopyright}
%\setcopyright{acmlicensed}
%\setcopyright{rightsretained}
%\setcopyright{usgov}
%\setcopyright{usgovmixed}
%\setcopyright{cagov}
%\setcopyright{cagovmixed}

% DOI
\doi{xxxxxx}

% ISBN
\isbn{978-1-4503-4263-6}

%Conference
\conferenceinfo{Sensys 2016}{November 14--16, 2016, Stanford, CA, USA}

\acmPrice{\$15.00}

%
% --- Author Metadata here ---
%\conferenceinfo{WOODSTOCK}{'97 El Paso, Texas USA}
%\CopyrightYear{2007} % Allows default copyright year (20XX) to be over-ridden - IF NEED BE.
%\crdata{0-12345-67-8/90/01}  % Allows default copyright data (0-89791-88-6/97/05) to be over-ridden - IF NEED BE.
% --- End of Author Metadata ---

\title{From Physical to Cyber: \\ Escalating Protection for Personalized Auto Insurance}

\numberofauthors{1}
\author{
\alignauthor 
        {Le Guan$^{\dagger}$, Jun Xu$^{\dagger}$, Shuai Wang$^{\dagger}$, Xinyu Xing$^{\dagger}$, \\Lin Lin$^{\dagger}$, Heqing Huang$^{\dagger}$, Peng Liu$^{\dagger}$, Wenke Lee$^{\ddagger}$}\\
        \affaddr{$^{\dagger}$The Pennsylvania State University, USA  ~~~~~~ $^{\ddagger}$Georgia Institute of Technology, USA}  \\
        \email{\normalsize \{lug14, jxx13, szw175, xxing, pliu\}@ist.psu.edu,
        \{llin, hhuang\}@psu.edu, wenke@cc.gatech.edu}
}

\CopyrightYear{2016} 
\setcopyright{acmcopyright}
\conferenceinfo{SenSys '16,}{November 14-16, 2016, Stanford, CA, USA}
\isbn{978-1-4503-4263-6/16/11}\acmPrice{\$15.00}
\doi{http://dx.doi.org/10.1145/2994551.2994573}

\maketitle
\begin{abstract}

Nowadays, auto insurance companies set personalized insurance rate based on data
gathered directly from their customers' cars.  In this paper, we show such a
personalized insurance mechanism -- wildly adopted by many auto insurance
companies -- is vulnerable to exploit. In particular, we demonstrate that an
adversary can leverage off-the-shelf hardware to manipulate the data to the
device that collects drivers' habits for insurance rate customization and obtain
a fraudulent insurance discount.  In response to this type of attack, we also
propose a defense mechanism that escalates the protection for insurers' data
collection. The main idea of this mechanism is to augment the insurer's data
collection device with the ability to gather unforgeable data acquired from the
physical world, and then leverage  these data to identify manipulated data
points.  Our defense mechanism leveraged a statistical model built on
unmanipulated data and is robust to manipulation methods that are not foreseen
previously. We have implemented this defense mechanism as a proof-of-concept
prototype and tested its effectiveness in the real world. Our evaluation shows
that our defense mechanism exhibits a false positive rate of 0.032 and a false
negative rate of 0.013.

\end{abstract}

%
% End generated code
%

%
%  Use this command to print the description
%

% We no longer use \terms command
%\terms{Theory}


\section{Introduction}
\label{sec:intro}

%% Background
Auto Insurers have long been known for pricing based on customers' evaluated
driving risks. Historically, drivers' risks were oftentimes determined simply
based on age, gender, model of the car and DMV records. With the recent
development in personalization algorithms and auto telematics, personalized
pricing strategy in auto insurance
industry has evolved into a new era. Consumers' insurance rate now could be
based on the actual driving data collected from their vehicles.

Led by Progressive Corporation, many auto insurance
companies offer their customers a voluntary discount
program, in which a customer needs to connect a telematics device to
his car through an \emph{On-Board Diagnostic-2 (OBD-2)} port. This device
records and sends driving data such as vehicle speed and \emph{Revolutions Per
Minute
(RPM)} to the insurers. Using these data,
insurers then calculate the dangerous driving behaviors
of the vehicle operator (e.g., the frequency of the hard brakes~\cite{progressiveterm}), and
analyze its measure to determine if the operator is eligible for an insurance
discount and the depth of the discount. According to a new market research
report~\cite{prnewswirereport}, the insurance telematics market size is expected
to grow from USD 857.2 Million in 2015 to USD 2.21 Billion in 2020, at a
compound annual growth rate of 20.9\%. While there are obvious
advantages in embracing personalized auto insurance for both businesses and end customers,
there also comes a new form of attack, in which
adversaries can exploit the algorithms underlying insurance personalization with the goal
of obtaining a fraudulent outcome.

In this paper, we show that miscreants can use off-the-shelf hardware to falsify the data
to insurance telematics devices, mislead insurance rate adjustment algorithms and
ultimately obtain financial profits to which they are not otherwise entitled.
Different from existing attacks against insurance telematics
devices~\cite{argusnews,forbesnews,191966}, one distinguishing feature of our new
attack is that it does not exploit any vulnerability resided in the devices.
Rather, it leverages services' own personalization mechanisms to alter insurers'
thought and decisions. As such, it indicates the current approaches to
cyber security are ill-equipped to address the vulnerabilities likely to exist
in all personalized auto insurance services.

%% Research Challenge, Goal and Idea
Since people can engage with technology in non-deterministic fashions, and there is no
standard reference to authentic user actions,
the key challenge in counteracting the aforementioned attack
is to distinguish if the data acquired from a user represent his authentic
actions or malicious manipulation. To
tackle the challenge, the paper further presents an effective anomaly detection
mechanism. The basic idea is to verify the legitimacy of the input data to underlying
personalization algorithms by using non-tamperable data acquired from the physical
world. More specifically, our detection mechanism uses unforgeable
acceleration measures to identify the vehicle speed manipulated by
miscreants.

%% Defense
We develop the aforementioned idea by first building a proof-of-concept system
prototype which emulates an insurance telematics device with a newly added embedded accelerometer.
The system prototype contains two components -- an OBD-2 reader and a
three-dimensional accelerometer. The OBD-2 reader is used for gathering
vehicle speed from a cyber space -- a car's OBD computer -- while
the accelerometer is for sensing vehicle acceleration from the physical world.
Although our system prototype does not prevent the physical access to the
accelerometer and OBD-2 reader, our work assumes the data acquired through
accelerometer are unforgeable for the simple reasons that auto insurers can easily
manufacture their insurance telematics devices with an embedded accelerometer and
protect it against the physical access by armoring it in a self-destruct box~\cite{mesh}.

Intuition might suggest our physical-space acceleration measures should
correlate with the variation in speed acquired from the aforementioned cyber
space. As a result, an instinctive reaction would be to compute vehicle velocity
from acceleration measures, and then use it for insurance rate customization.
However,  this is infeasible. First, insurance rate customization needs accurate
velocity measure and, in practice, existing techniques~\cite{virginiareport, 
Yoon:2007:SST:1247660.1247686, Mohan:2008:NUM:1460412.1460450} do not provide
adequate accuracy in velocity measurement. For example, prior 
studies~\cite{GoogleTechTalks, Woodman07anintroduction} have
indicated  that the accelerometer-based velocity estimation adopted in Inertial
Navigation  System (INS) suffers from integration drift -- small errors in the
measurement of acceleration and angular velocity are integrated into
progressively larger errors in velocity\footnote{In~\cite{Woodman07anintroduction}, 
a quantitative study indicates that the average error
in position grows to over 150 meters after 60 seconds of operations.}.
Second, a car's onboard diagnostic
computer can accurately measure the rate of vehicle speed using the rate of
rotation of a drive shaft, whereas an accelerometer is susceptible to noise
caused by poor road condition and engine vibrations etc. As a result, the
measures of speed variation -- from cyber and physical spaces -- typically
exhibit unobserved heterogeneity.

As part of our detection mechanism, we address the challenge of unobserved
heterogeneity using a statistical model. In particular, the statistical model
follows the framework of mixture regression models, which captures unmeasurable
noise in the physical world, and reflects the
relationship between physical-space acceleration and the
variation in speed acquired from the cyber space. For each physical-space
measure, our statistical model outputs a \preint.
% prediction interval.
In detecting speed manipulation, we examine if the speed variation measure from
cyber space extends outside our \preint.  Once
discovering the cyber-space measure falls outside the \preint, our detection
mechanism flags the measure as an anomaly.

Our detection mechanism provides several advantages.
Most notably, it escalates protection for personalized insurance program.
With
our detection mechanism, auto insurers can identify the unlawful activities of
their policyholders, and prevent losses from fraud. In addition, our mechanism is
robust to previously unseen manipulation strategies. We do
not construct the underlying statistical model specific for certain manipulation
strategies. Rather, we use unmanipulated data to model relationship between
cyber-space speed variations and physical-space acceleration, and then utilize
the model to evaluate the speed measures acquired from untrustworthy cyber space.

In summary, the paper makes the following contributions.

\begin{itemize}
  \item We describe a generic attack against personalized car insurance program that
    allows miscreants to alter their unwanted driving behaviors and
    potentially obtain unlawful financial profits.
  \item We demonstrate the new attack against 7 different insurance
    telematics devices, disclose our findings to corresponding auto
    insurance providers and raise their attention to the new security problem.
  \item We present a noise-resilient detection algorithm and demonstrate how to
    use it to counteract the manipulation of driving behaviors.
  \item We study the effectiveness of our detection algorithm using a
    1034-mile driving trace, and show our algorithm achieves a false
    negative rate of 0.013 and a false positive rate of 0.032.
\end{itemize}

The rest of the paper is organized as follows. Section~\ref{sec:related-work}
surveys related work. Section~\ref{sec:attack}
describes and demonstrates an attack against insurance telematics device.
Section~\ref{sec:defenseoverview} presents the overview of our defense system.
Section~\ref{sec:proto} shows a proof-of-concept defense prototype that emulates
an insurance telematics device with an accelerometer embedded.
Section~\ref{sec:model} elaborates a statistical model that identifies the
manipulation of vehicle speed using the data gathered from the defense prototype.
In Section~\ref{sec:eval}, we evaluate our
detection mechanism, followed by technical discussion in 
Section~\ref{sec:discussion}. Finally, we conclude the work in 
Section~\ref{sec:conclusion}.

\section{Related Work}
\label{sec:related-work}

There are three lines of work most closely related to ours -- cyber attacks in automotive
contexts, intelligent transportation system, and anomaly detection. In this section, we discuss these work in
turn.

\vspace{3pt}
\noindent \textbf{Attacks in Automotive Contexts.} In recent years, a
significant amount of research has been performed in the context of automotive
systems. Research in 
this domain mainly focuses on two aspects -- attacks against vehicle anti-theft
systems~\cite{Francillon11, 193260} and attacks against auto OBD
computer~\cite{191966, koscher2010experimental, snapshotattack,
miller2013adventures}. 

Attacks against vehicle
anti-theft systems exploit the software and hardware weakness of smart car
keys. For example,
Verdult et al. reverse-engineered a passive RFID tag embedded in electronic
vehicle immobilizers and disclosed the weakness in its design and 
implementation~\cite{193260}. In another research work, Francillon et al.
analyzed the protocol of immobilizers used in modern cars, and demonstrated an
adversary can enter and start a car by relaying messages between the car and
smart key~\cite{Francillon11}. 

Attacks against auto OBD computer target the internal
network of automotive systems. As all modern automobiles rely on a broadcast
network to connect the various components of a car -- including the engine,
transmission, brakes, airbags, lights, and locks -- quite a bit of previous research
emphasizes on discovering and analyzing the security flaws of the broadcast
network. In~\cite{koscher2010experimental, Checkoway:2011:CEA:2028067.2028073},  
Koscher et al. systematically analyzed the fragility of auto broadcast network.
In follow-on work~\cite{191966, foster}, Foster et al. demonstrated that an attacker 
can compromise a telematics control unit, connect to a vehicle and take the
control of the vehicle remotely. 

While our attack is performed in the same context, it is significantly distinct.
In terms of the goal of our attack, an adversary aims to obtain a fraudulent
insurance discount rather than taking the control or gaining access to a vehicle.
From the technical perspective, the success of our attack does not
rely upon any vulnerabilities or flaws resided in the software or hardware of an
automotive system. In fact, our attack only distorts or counterfeits the data
gathered from automotive systems. 

\vspace{3pt}
\noindent \textbf{Intelligent Transportation System.}
Intelligent transportation system~\cite{intelligenttransportation}
aggregates sensory data from multiple sources for different purposes, such as
realtime traffic delay prediction~\cite{vtrack, claudel2008guaranteed},
road and traffic condition monitoring~\cite{Mohan:2008:NUM:1460412.1460450,pothole},
and driving behavior estimation~\cite{estimatingbehavior,chensecon15,Paefgen:2012:DBA:2406367.2406412}, etc.
Among these applications, many use physical-world data to predict
specific events. To name a few, 
NeriCell~\cite{Mohan:2008:NUM:1460412.1460450} uses sensing components
on smartphones to detect potholes, bumps, braking, and honking, and thus obtain
a picture of road surface quality as well as traffic conditions.
Nericell addresses the problem of accelerometer re-orientation when the phone's orientation
changes over time.
Similarly, Pothole Patrol~\cite{pothole} is designed to identify potholes using accelerometer data.
It avoids the re-orientation problem by mounting the accelerometer at a known orientation.
In our work, our detection model does not rely
on orientation. Specifically, our model takes input from a 3-axis accelerometer, and
predicts a velocity variation range in a direct manner.

\vspace{3pt}
\noindent \textbf{Anomaly Detection.}
In terms of defense, our work resembles anomaly detection which identifies
patterns in data that do not conform to an expected behavior. In the past
decades, anomaly detection has been studied in several research communities
across a large number of data domains, including high-dimensional
data~\cite{Aggarwal:2001:ODH:375663.375668}, uncertain
data~\cite{Aggarwal08outlierdetection}, streaming
data~\cite{Aggarwal12eventdetection,Aggarwal05}, network 
data~\cite{Gao:2010:COE:1835804.1835907,Gupta:2012:ICM:2339530.2339667} and time
series data~\cite{Ma:2003:OND:956750.956828}. Though the nature of the data in
our application 
fits the data type in time series, applying the techniques developed for that
data domain is not straightforward, for the reason that the exact notion of an anomaly is
different for different application domains. For example, in the stock market
domain fluctuations in the stock value might be normal while similar deviation
in our application domain is considered as an anomaly. In addition, the data in our
application contains a lot of noise that tends to be similar to the actual
anomalies. Consequently, anomaly detection techniques that design specifically for
other applications are unlikely to be effective against our attack.

\section{Attack against Personalized Insurance}
\label{sec:attack}

In this section, we show that insurance telematics devices are vulnerable to a new
class of attack
in which an adversary can manipulate the data sent to the devices with the goal of
obtaining a fraudulent discount from insurers' personalized insurance
program.
In particular, we first describe the overview of insurance telematics devices followed
by technical background relevant to the devices. Then,
we reverse engineer the devices and explore how they operate. Based
on the knowledge acquired through our reverse engineering, we
demonstrate potential attacks against insurers' personalized insurance
program. Last but not least, we discuss the cost of our attack and some ethical
concerns.

\subsection{Overview of Telematics Device}

An insurance telematics device is typically used for monitoring the driving
habits of auto insurance policyholders. It collects information about when a
policyholder uses his vehicle, how far he drives and whether he drives with
sharp braking. In addition, it collects the Vehicle Identification Number (VIN).
Using these information together, an insurer calculates personalized insurance
rates for its customers. In general, policyholders who volunteer to install a
telematics device can get a markdown on their auto insurance of up to 50\% if
they minimize hard braking, the time behind the wheel and the number of minutes
they spend driving during higher risk hours~\cite{progressiveterm, righttrack,
drivewisereview, indrive}.

\begin{figure}
\centering
  \includegraphics[width=0.8\columnwidth]{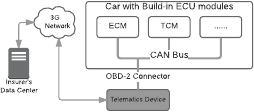}
  \caption{The overview of an insurance telematics device.}
  \label{fig:telematicsarch}
  %\vspace{-6mm}
\end{figure}

Typically, a telematics device runs a micro-kernel on an embedded processor.
Using a cellular modem embedded, it communicates with an insurer's data center
and sends back to the insurer the aforementioned
information~\cite{bauer2012monitoring, snapshotattack}. The telematics device
physically connects to a car's OBD-2 port -- SAE J1962 connector located under
the driver's side dashboard -- so that it can send specially formatted
diagnostic command messages over CAN bus, communicate with many Electronic
Control Unit (ECU) modules and monitor the vehicle operation. To be eligible to
an insurance discount, a policyholder must plug in the device and drive with it.
The telematics device periodically pings an insurer's data center and ensures it
is not physically disconnected. Figure~\ref{fig:telematicsarch} illustrates how
a telematics device communicates with an insurer's data center and interacts
with ECU modules built into the car. 
% To facilitate policyholders to keep track
% of their driving activities, insurers provide them with a web interface that
% shows the number of hard brakes they incur, the number of miles they drive and
% the amount of time in each trip.

\subsection{Technical Background}

% As is mentioned above, a telematics device is physically attached to a car's
% OBD-2 port, and communicates with many ECU modules through CAN bus. Here, 
In this section, we describe
the technical background of ECU modules, CAN bus, and OBD-2.

\subsubsection{ECU Modules}

Electronic Control Unit (ECU) is a generic term for a device that controls one or more
electrical systems in a vehicle. It is the ``brain" of a motor vehicle that
monitors and controls vehicle operation such as brake, electronic fuel injection
and ignition timing etc. Some typical ECU modules include Brake Control Module
(BCM), Transmission Control Module (TCM) and Engine Control Module (ECM) etc.
As is shown in Figure~\ref{fig:telematicsarch}, ECU modules can communicate with
each other through Controller Area Network (CAN bus) without a central computer.

\subsubsection{CAN Bus}

CAN bus is a multi-master broadcast serial bus standard designed to allow many
ECU modules to communicate with each other within a vehicle. When an ECU module
sends a message, every other ECU module on the bus receives it and can choose to
respond to it or ignore it. 
% Every message has a unique priority. The higher the
% priority is, the more likely it will be sent first. For example, brake may have
% the highest priority. 
% When a vehicle operator apply brake he definitely wants
% the car to slow down as soon as it can. 
CAN defines the structure and the way
data are transferred between ECU modules.

\subsubsection{OBD-2}

OBD-2 is a standard that specifies the type of diagnostic connector and its
pinout, the electrical signaling protocols available, and the messaging format.
An OBD-2 port is an physical interface, through which an OBD-2 device (e.g., an
insurance telematics device) can access to various ECU modules and retrieve
their statuses such as engine temperature and engine speed etc. As is
illustrated in Figure~\ref{fig:telematicsarch}, an OBD-2 device sends specially
formatted diagnostic command messages over the CAN bus. ECU modules on the
network send out the requested status information over CAN when asked. To query
vehicle speed, for example, a technician can enter Parameter ID (PID)
\texttt{0x0D} into an OBD-2 device which then sends the corresponding message
over the CAN bus. The ECU module that knows the vehicle speed returns the
vehicle speed.

\subsection{Reverse Engineering}
\label{sec:understanding}

%% Goal
In order to launch an attack against insurers' personalized insurance
program, we reverse engineer their telematics devices\footnote{
  There has been similar work from the security community to cheat for
  30\% discount on insurance~\cite{communityinsurance}. We figured it out independently, 
  made it more sophisticated, and designed a kit to facilitate such attacks.}.
We are interested in a number of questions: In what condition, and
at what frequency are information collected? When does a telematics device
start and end data collection? Have insurers deployed and used any anomaly
detection mechanisms to prevent potentially malicious data manipulation?

%% Methodology
To answer the aforementioned questions, we built a testbed that
simulates vehicle ECU modules. The testbed allows us to monitor the request
messages of a telematics device over the CAN bus so that we can better understand how
the device works and observe what information it collects. In addition, the
testbed allows us to control the response to the request. 
% Using it along with
% insurers' web interfaces, we therefore can examine if insurers have already
% deployed any anomaly detection mechanisms to prevent unlawful data manipulation.

%% Testbed
In establishing our testbed, we used a commercial off-the-shelf (COTS)
hardware -- ECUsim 2000
professional firmware edition~\cite{ecusim} -- to simulate ECU modules built in a car.
The COTS hardware simulator provides physical connection to OBD-2
devices through
a standard SAE J1962 female connector. Through an USB connection to the
hardware simulator, we can enable software control to it, monitor OBD-2 request
messages over the CAN bus and control the response to the request messages.

%% Testing Testbed Setup
% We tested the setup of our testbed. More specifically, 
We connected
7 different insurance telematics devices to the ECU simulator,
and then used a C program to operate the simulator. In this
setup, we passively monitored the communication between the simulator and those
devices, but failed to observe any messages. To understand the root
cause, we inspected the electronic system of a motor vehicle and compared it with
that of our ECU simulator. In particular, we used an oscilloscope to observe the
voltage changes of the SAE J1962 female connector on both systems. According to
CAN specification, we measured
the change of the electrical signal at PIN 6, PIN 14 and PIN 16\footnote{A
standard SAE J1962 connector contains 16 pinouts in which PIN6, PIN 14 and PIN
16 is defined as CAN-High signal, CAN-Low signal and battery voltage (VCC)
respectively.}.

%% Findings and Adjustment
We observed the same patterns on voltage change from the motor vehicle and our
simulator when the car key is not beyond the `ignition on' position.
As the motor vehicle starts,
we found the vehicle system yields a voltage jump of $1 \sim 2$ volt at PIN 16,
whereas we did not observe such a jump from our simulator. Our hypothesis is
that our simulator does not emulate a car's ignition and insurance telematics
devices
start to request information from a motor vehicle only when they detect such
a voltage jump. Therefore, we utilized an adjustable voltage power supply to
power the simulator. In particular, we tuned up the voltage volume at PIN 16
from 12V to 13.3V. From our simulator, we observed all the tested telematics
devices
start to request information about vehicle operation after switching their
operation voltage to 13.3V.

\begin{figure*}
\centering
  \includegraphics[width=1\textwidth]{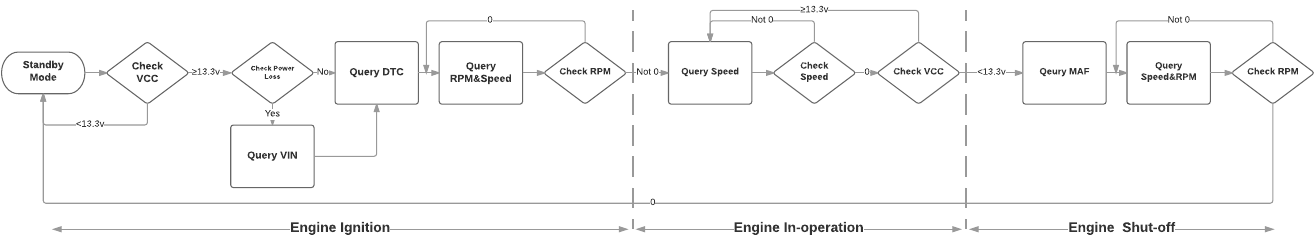}
  \caption{The workflow of a telematics device in general.}
  \label{fig:snapshotprotocol}
  %\vspace{-4mm}
\end{figure*}

%% Summarizing findings from Reverse Engineering
We set up our testbed with our ECU simulator and an adjustable voltage power
supply, and used it to understand the workflows of
telematics devices from 7 auto insurance providers in the US. In particular,
we used our testbed to monitor the request messages from each telematics device, and
vary the response accordingly. 
In this work, we disclose the workflows of these telematics devices in general.
% Due to page limit, we only describe the workflow
% of Progressive's Snapshot~\cite{progressivesnapshot}, the most popular insurance telematics
% device in the US. Telematics devices from other insurers follow the similar
% operation.

We mark off the workflow of a telematics device into three phases -- \emph{engine
ignition}, \emph{engine in operation} and \emph{engine shut-off}.
Figure~\ref{fig:snapshotprotocol} illustrates the workflow of a
telematics device in each phase.

%% Findings in "ignition on" status
\noindent{\textbf{Engine Ignition.}}
As is discussed above, the operation voltage of a telematics device is 13.3V.
When detecting 13.3V, the telematics device starts to request information about vehicle
operation including the speed and engine Revolutions Per Minute (RPM).
Using RPM, the telematics device determines if the
ignition is on. In particular, when RPM reading is above zero, the telematics
device sets its internal state to ``ignition on".

As is shown in Figure~\ref{fig:snapshotprotocol}, every time upon its physical
contact with the car,
the telematics device also requests Vehicle Identification Number (VIN). This
indicates that the insurer might be aware of tampering if a policyholder attempts to
earn an unlawful discount by unplugging his telematics device right before a trip
and plugging it back after the trip.

In addition to RPM, speed and VIN, we also surprisingly found that some telematics
devices attempt to access Diagnostic Trouble Codes (DTCs) of the motor vehicle,
which indicate the malfunctions within the vehicle. While the corresponding insurers claim
they calculate a discount only based on a variety of factors related to driving
activities, considering the malfunction information may be used a factor to
affect one's insurance rate and the insurers fail to specify the collection of
such information in their privacy policy, we believe it is suspicious and
inappropriate to collect malfunction information within a car without a clear
declaration.

%% Findings after "ignition on"
\noindent{\textbf{Engine in Operation.}}
Once the engine is started (i.e., the operation voltage reaches 13.3V, and RPM
and MAF readings are above zero), the telematics device starts to query vehicle
speed repeatedly. In particular, the device sends the vehicle speed request over
the CAN bus once every second presumably because most motor vehicles refresh
their OBD readings at about the same frequency. As the vehicle is moving, the
telematics device also monitors abrupt decreases in vehicle speed. More
specifically, the telematics device beeps if it discovers a change more than certain
kilometers\footnote{Different insurers set different thresholds. 
For example, Progressive considers velocity decrease of more than 7 mile/h 
in a second to be dangerous.} in consecutive speed readings.

%% Findings after "ignition off"
\noindent{\textbf{Engine Shut-off.}}
The telematics device stops querying vehicle speed when its operation voltage
drops below 13.3V. However, it does not terminate completely. As shown in
Figure~\ref{fig:snapshotprotocol}, the telematics device instead checks the speed
of the engine and the mass of air flowing into the engine presumably because
some cars shut off their engines at stoplights for fuel saving and emission
reducing and the insurer does not want to mistakenly count such engine restart
as the beginning of a new trip. In fact, the telematics device marks the end of a
trip only when its operation voltage is below 13.3V, and the readings of MAF and
RPM are equal to zero.

\begin{figure}[h]
\centering
  \includegraphics[width=0.9\columnwidth]{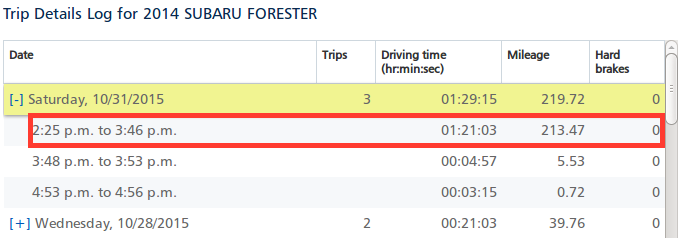}
  \caption{An evidence showing the absence of anomaly detection on the insurer's server.}
  \label{fig:abnormaldetection}
  %\vspace{-2mm}
\end{figure}

In addition to exploring the workflows of the insurance telematics devices, we utilized our
testbed to examine if insurers have already deployed any anomaly detection
systems to exclude obvious anomalies, e.g., extremely high vehicle speed. In
particular, we emulated a trip where our testbed constantly responded
telematics devices with a vehicle speed of 157 miles per hour for
about 80 minutes. Then, we observed the trip through the aforementioned
insurers' web interfaces. To our surprise, insurers involved in our experiment
all log and present our
emulated trip on their web interfaces (see Figure~\ref{fig:abnormaldetection}).
As insurers log legitimate trips and present corresponding information through
their web interfaces, the presentation of our emulated trip indicates the absence
of anomaly detection on insurer side.

\subsection{Attack}
\label{sec:2attack}

As is discussed earlier,
an insurer calculates a personalized insurance rate for each policyholder
based on the data gathered from his telematics device, and the factors that
affect one's insurance rate are publicly known~\cite{progressiveterm, drivewisereview}. This provides an unlawful
customer with a possibility of altering the data to his telematics device and
obtaining a fraudulent discount. Here, we propose two attacks against
personalized insurance program provided by 7 different insurers and demonstrate both attacks could potentially
allow an adversary to earn an unlawful discount.

\subsubsection{Offline Attack}

Our first attack is an offline attack which roots in conventional replay
attacks. In particular, an adversary first records a data trace representing
safe driving activities. Then, he replays the data trace to his telematics device.
Since the data trace represents safe driving activities, the adversary can
mislead an insurer into believing he is eligible to an insurance discount even
though he is a high-risk driver.

To demonstrate this attack, we first developed a data collection platform which
allows us to obtain the driving information from a car's OBD
computer. The data collection platform consists of two components: a bluetooth
OBD-2 adapter physically connected to a car's OBD-2 port, and an Android phone
that communicates with the car's OBD computer through the
bluetooth adapter.

Using the data collection platform, we recorded a 10-mile trip for 15 minutes.
Then, we replayed the trip by responding telematics devices with corresponding
data we collected. In particular, we connected telematics devices to the testbed
we discussed in the previous section, and then used it to respond those devices
accordingly. We performed this replay attack once every day in a week-long
period. Through those insurers' web interfaces, we observed all insurers logged our
replayed trip for seven consecutive days. Again, this indicates these insurers have
not yet deployed any anomaly detection system to identify obvious anomalies
(i.e., exactly the same driving activities shown to insurers for many days).

\subsubsection{Online Attack}

Despite effectiveness, the replay attack can be easily detected. As the replay
attack does not need to be carried out in a moving vehicle -- and some
insurance telematics devices have been designed with a GPS component -- one instinctive
detection scheme is to use GPS to collect vehicle location and examine if the
car is in motion. In response to the limitation of replay attacks, we propose
another attack.

Our second attack is an online attack which manipulates the data representing
dangerous driving activities. Different from the first attack, this attack is
carried out on the fly. In particular, we built a man-in-the-middle (MITM) box
which bridges a car's OBD-2 port and a telematics device. Our MITM box monitors
the response messages to a telematics device. Once it detects a response message
indicating an abrupt decrease in vehicle speed, it will flatten speed changes
because a sudden drop in speed will be transmitted back to insurers and has a
negative impact upon the policyholder's discount. 
With the knowledge of the insurer's threshold that identifies a hard brake, our
MITM box simply responds the highest value that is below the threshold.
Since our MITM box is installed
inside a motor vehicle carrying out data manipulation while the vehicle is in
motion, this online attack is resistant to the aforementioned defense which
utilizes location information to identify the malicious attack.

\begin{figure}[t]
  \centering
  \includegraphics[width=0.9\columnwidth]{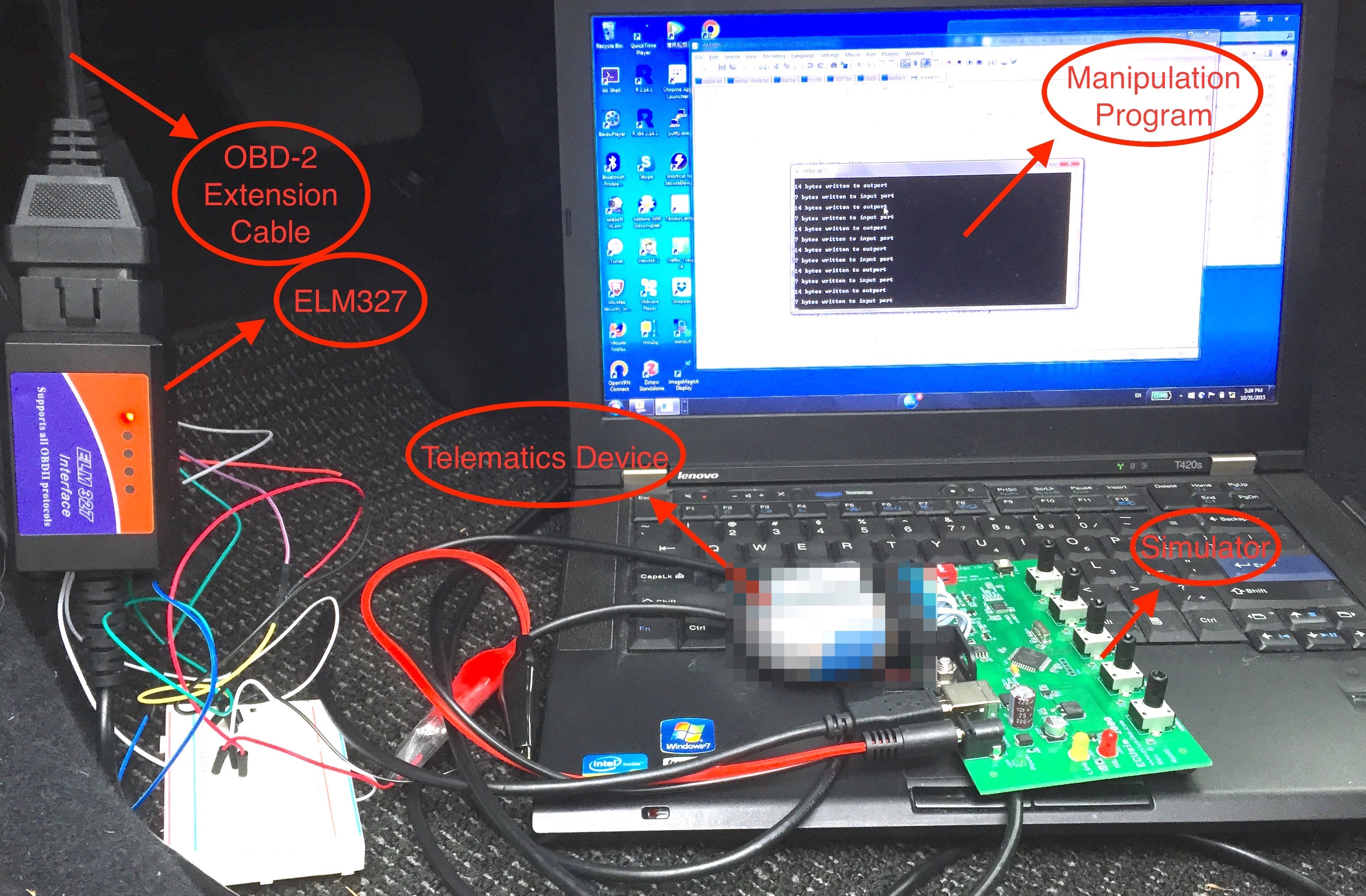}
  \caption{The man-in-the-middle box that bridges a car's OBD-2 port and a telematics device.}
  \label{fig:mitm}
  %\vspace{-6mm}
\end{figure}

We implemented an MITM box prototype
which consists of three components: (1) an ELM327 adaptor that connects a
laptop and a car's OBD-2 port, (2) a C program that monitors response messages
from a car's OBD computer, and
manipulates them if necessary and (3) an ECUsim 2000 simulator that
encodes forwarded OBD-2 readings (potentially manipulated)
into CAN message.
Figure~\ref{fig:mitm} shows how these components are assembled.
We installed our MITM box in a 2014 Subaru Forester
and experimented its effectiveness in a trip.
In our experiment, we intentionally
pushed brake pedal hard. However, we did not observe those telematics devices
capture such dangerous driving activities.
%  This indicates the success of our
% attack.

\subsection{Discussion}

\subsubsection{Cost}

As is described above, we demonstrated our attacks using off-the-shelf hardware
(i.e., ECU Sim 2000 and ELM 327 cable). These hardware devices typically cost
a couple of hundred dollars. However, the investment in these devices does
not frighten miscreants off, especially considering the long-term insurance
discounts to which they are not otherwise entitled. In addition, following
Moore's Law, we can expect the fall in hardware price and thus the low cost of
our attack.

\subsubsection{Ethical Concerns}

In this work, our goal is not to obtain unlawful profits. Rather, we expose the
vulnerability resided in insurance telematics devices. As such, we purchased a
car insurance coverage from 7 different insurers, enrolled our own vehicle in a
discount program and experimented our attacks in a time window of three
weeks. Before and after our 3-week experiment, we did not plug the telematics
devices to our registered vehicle. Thus, this allows us to experiment attacks
without jeopardizing insurers' business. Note that once policyholders enroll in
a discount program and have their telematics devices plugged into the
registered vehicle for 30 days, insurers evaluate their driving behavior and
apply their initial discount.

% We have disclosed our understanding of the problems to 7 insurers whose devices
% are involved in our experiments. As of submission, with exception of
% Progressive~\cite{progressivesnapshot} and Liberty Mutual~\cite{righttrack}, we
% have not yet received permission to disclose our findings on other insurers nor
% to reveal their names publicly. Progressive and Liberty Mutual were supportive
% of our work, appreciative that we had informed them in advance, and intimated
% that they will take immediate action on the new security threat.

\section{Defense Overview}
\label{sec:defenseoverview}

The aforementioned attacks exploit the fact that the auto insurance discount
program is
lack of an effective mechanism to check the integrity of the data acquired from
end users. Once a telematics device is handed out to the end users, an insurer
loses their control to the data collection. As a result, unlawful customers can
manipulate their telematics devices in various ways with the goal of obtaining fraudulent
discounts. In other words, without an integrity check to the data collected from
users, the auto insurance discount program is highly vulnerable to information
manipulation.

In the past, many approaches have been designed to protect data integrity. Most
notably, traditional computer systems utilize digital certificate to protect data
integrity in transit. However, such approaches cannot be easily applied to many
complex, highly-customized systems due to the difference in system design
principle. Different from conventional computer systems, an auto system for example
is not designed
with a component to manage the risks to data availability, confidentiality and
authenticity.

In fact, it is impractical to augment complicated and highly customized system
with the capacity of managing threats to data. Again, take the auto system for
example. From the perspective of insurance companies, it would be unrealistic if
their data collection relies on a fundamental system change to their customers'
motor vehicles.

In this work, we propose a new mechanism that can reduce auto insurers' exposure to
data integrity threats. This new mechanism does not require any modification to
auto systems. The basic idea of this mechanism is to (1) on the client side, extend telematics devices
to collect unforgeable data and (2) on the server side, augment insurers' data centers with the
ability to use the data to minimize the threat to aforementioned data
manipulation. More specifically, we introduce a defense mechanism that contains
two components -- a proof-of-concept system prototype that emulates an
insurance telematics device with a tamper-resistant accelerometer, and an effective
detection algorithm identifying data manipulation on a remote server.
% In the following, we first present the brief overview of our defense mechanism.
% Then, we discuss the trust model underlying our defense mechanism, followed by
% a threat model.
The detail of our defense mechanism is discussed in Section~\ref{sec:proto}
and~\ref{sec:model}, which describe the client and server side implementation respectively.

\subsection{Defense Mechanism Overview}
\label{sec:systemmodel}

As is mentioned above, our defense mechanism contains two components. One
component is for data collection and the other is for manipulation detection.

The data collection component is equipped on a motor vehicle. It gathers two
types of data -- vehicle speed acquired from
a car's OBD computer and accelerations collected through
tamper-resistant physical sensors. The data collection component periodically
sends the collected data back to a central server.

The manipulation detection component runs on the central server. It processes the
received data and then uses a statistical model to identify the data potentially
manipulated. In particular,
the manipulation detection component processes unforgeable accelerations as well
as the vehicle speed which might be potentially manipulated. Using a statistical
model, the detection component takes the input of accelerations and outputs a
\preint. The \preint represents a normal region. In
identifying potential manipulation, the detection component examines if
the speed variation derived from the consecutive speed readings acquired from auto
system lies in the normal region.

\subsection{Trust Model}

% As is discussed in the previous section, an adversary can easily tamper the data to
% a telematics device that collects user driving behaviors for personalized
% insurance pricing. Therefore, 
Our defense
mechanism assumes that the data acquired from a car's OBD
computer is untrusted. As an insurer itself is an honest party responsible for detecting
potential fraud, our defense mechanism trusts the data collected through
sensors built in the telematics device. In addition, our defense mechanism trusts
the communication channel between the telematics device and the insurer's server.
% Figure~\ref{fig:trustmodel} illustrates the overall trust model of
% our proposed defense system.
We discuss the feasibility of our assumptions below.

%\vspace{3pt}
\noindent\textbf{Tamper-resistant Chip.} To obtain unforgeable
accelerations, an insurer can incorporate an accelerometer to its telematics
device and then safeguard the device against physical
attacks\footnote{\cite{snapshotattack} demonstrates an attack that physically
accesses to the chipset on a Progressive's Snapshot device and intrudes in the
on-board network of a vehicle.}. This can be achieved by protecting a telematics
device with tamper-resistant techniques~\cite{tamperproof}. For example, an
insurer can build its telematics devices using sensor mesh that constantly
monitors any interruption or short-circuit on the chip~\cite{mesh}.
Alternatively, ARM TrustZone extension can be used to attest the integrity of
the sensors~\cite{sensorintegrity}. In fact, any device that conforms to
\emph{Federal Information Processing Standard (FIPS 140-2) [FIPS]} Level 4 is
required to provide a hard, opaque removal-resistant coating with hardness and
adhesion characteristics such that attempting to peel or pry the coating will
have a high probability of resulting in serious damage to the device.

%\vspace{3pt}
\noindent\textbf{Trusted Transmission.}
In transmitting data to the insurer's server, a secure communication channel can
be easily established. It has been disclosed in~\cite{snapshotattack} that, since some telematics devices
transmit data through a 3G communication channel in plaintext using FTP protocol
-- and the devices typically do not utilize any network authentication mechanisms --
an attacker could easily set up a faked cellular phone base station
and launch a man-in-the-middle attack.
Following careful implementation and design, these vulnerabilities can be easily prevented.
In our defense mechanism, each telematics device should be embedded with a digital
certificate, and perform an SSL authentication each time a communication channel is established.

\begin{figure}
\centering
  \includegraphics[width=.8\columnwidth]{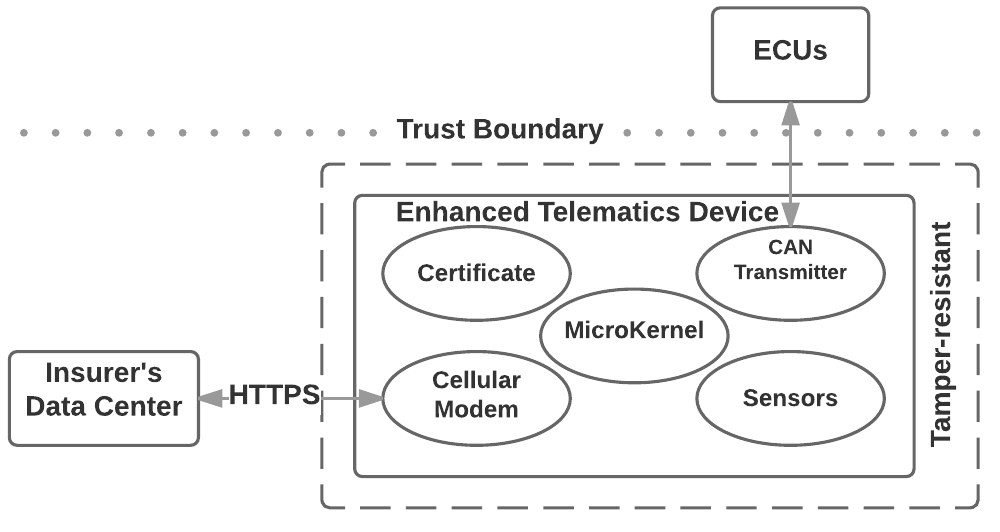}
  \caption{The trust model underlying the defense mechanism.}
  \label{fig:trustmodel}
  \vspace{-1mm}
\end{figure}

\subsection{Threat Model}

The main goal of our defense mechanism is to detect the sophisticated attack
discussed in the previous section. In fact, there are many other possible
threats to insurance telematics devices. 
First, an unlawful customer might equip his telematics device into a low-risk driver's
vehicle.
Second, he/she could also connect his/her telematics device to an equipment that
supplies electronic power and responds the device with data representing a
car in parking status.
Lastly, he/she 
may unplug the device before his/her trip and then plug it back after.
Although these threats could
potentially cause economic loss for insurers, they are out of our defense
umbrella because they could be easily counteracted. 

\section{Defense Prototype}
\label{sec:proto}

% In this section, we describe a proof-of-concept defense prototype on the client
% side. 
Our prototype on the client
side emulates an insurance telematics device with an embedded
accelerometer. It is composed of an OBD-2 I2C
adaptor~\cite{obd2i2c}, a microcontroller board based on the
ATmega2560~\cite{arduino} and a GSM/GPRS board~\cite{gsmgprs}.

The OBD-2 adaptor connects to the auto ODB-2 interface and gathers
speed readings of a vehicle through SAE J1962 connector. Using a built-in MEMS
based accelerometer, it also measures the acceleration of the vehicle. The
microcontroller board retrieves acceleration and speed readings through the
OBD-2 adaptor and temporarily logs them to a microSD card. Through the GSM/GPRS
board, the microcontroller sends the acceleration and speed readings to a remote
server, where our anomaly detection algorithm is running to examine the
legitimacy of the data.

\section{Defense Algorithm}
\label{sec:model}

\begin{figure*}[t]
\centering
\includegraphics[width=0.95\textwidth]{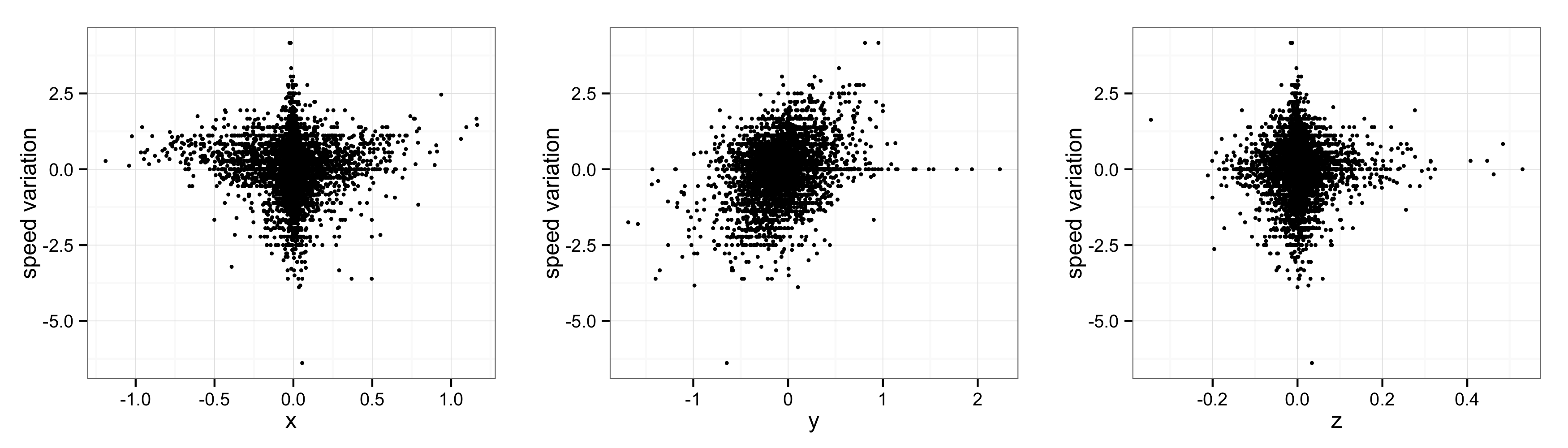}
\caption{The scatter plots of the speed variation against the acceleration
measures in each dimension in Euclidean space.
\label{fig:scatteryx} }
%\vspace{-4mm}
\end{figure*}

In this section, we present a statistical model that can be incorporated to the
aforementioned defense mechanism and counteracts the attack discussed in
Section~\ref{sec:attack}. As is discussed in Section~\ref{sec:defenseoverview},
the basic idea of our countermeasure is to use tamper-resistant acceleration
measures to evaluate the legitimacy of the speed data acquired through an OBD2
port. Developing this idea, we introduce a statistical approach to model the
relationship between speed variation and acceleration measure. Using this model,
we then examine whether a newly observed speed measure is manipulated.

%%The design principle of our statistical model is based on the observation of the
%%data collected through our aforementioned system prototype.
Figure~\ref{fig:scatteryx} illustrates three scatter plots, each of which
indicates the variation in vehicle speed against the
acceleration measure in each dimension. 
Since acceleration reflects the change in speed,
intuition suggests that the speed variation
and acceleration should follow a trend. In contrast, we observe that the
acceleration measures in each dimension are spread out across the speed
variation, and there is no clear trend between them. In statistics,
this is called {\em unobserved heterogeneity} which is typically contributed by
unobservable covariates (i.e., unmeasurable factors). For example, poor road
conditions can be an unmeasurable factor, resulting in acceleration drift which
contributes to the mismatch observed in Figure~\ref{fig:scatteryx}. 

% \comment{
% In addition, the unobserved heterogeneity occurs when some grouping structures
% reside in the acquired dataset. For example, the relationship between
% acceleration and speed measures is different if one group of drivers mount
% accelerometer in the horizontal position while the other group place it in
% vertical.
% }

As is discussed earlier, we need a model to capture the relationship between the
acceleration measure and speed variation (i.e., $y=f(x)$ where $x$ and $y$ represent an
acceleration measure and a velocity variation respectively). Here, an instinctive  approach
is to use a single regression model to fit a function for both measures.
However, a single regression approach typically suffers from the unobserved
heterogeneity problem, and we do not want the aforementioned unobserved
heterogeneity to influence the regression model accuracy. Therefore, we develop
our statistical approach in the framework of mixture regression models which
typically provide better control for unobserved heterogeneity.

In this section, we present how to model the speed variation and acceleration
measure using mixture  regression models. More specifically, we begin with the
introduction to the framework of finite mixtures of Gaussian regression models.
Then, we discuss how to use this framework to model the relationship between the
acceleration and speed variation. Finally, we present how to perform
manipulation detection using our model.

%%\subsection{Technical Background}
\subsection{Framework of Mixtures of Gaussian Regression Models}
\label{sec:background}

% \comment{
% Statistical mixture models have long been used for
% distribution density estimation~\cite{doi:10.1080/01621459.1995.10476550,
% doi:10.1080/10618600.1998.10474772}. The objective of the modeling analysis is
% to define the probability distribution model and evaluate inferences over the
% model parameters based on the fit to a specific data set. 
% }

Applications of mixture
models have appeared in various fields including biomedical
studies~\cite{PMID:23629459, Lin03062015, 10.1371/journal.pcbi.1003130},
economics~\cite{RePEc:ecm:emetrp:v:57:y:1989:i:2:p:357-84, 10.2307/2999563} and
marketing research~\cite{10.2307/3172759, 10.2307/30038851}.
Within the family of mixture models, the mixture of Gaussian
regression models has the tight structure of a parametric model and
retains the flexibility of a nonparametric method. As such, it provides 
better control for unobserved heterogeneity. In this work, we develop our
statistical model in the framework of the mixtures of Gaussian regression
models. Here, we describe this framework as follows.

Notationally, we say $y$ depends on the vector of covariates $x$
in a mixture of $J$ different ways if
\begin{align} \label{eq:mix}
y\sim \sum_{j = 1}^J \pi_j N(y|x' \beta_j, \sigma^2_j),
\end{align}
where $\pi_1, \pi_2, \cdots, \pi_J$ are probabilities with a sum of 1. $N(x' \beta_j, \sigma^2_j)$
denotes the density of a Gaussian distribution with mean $x' \beta_j$
and variance $\sigma^2_j$. $\beta_j$ is the regression
coefficient for the $j^{th}$ covariate $x_j$. All parameters $\{\pi_{1:J},
\beta_{1:J}, \sigma^2_{1:J}\}$ construct a parameter set denoted by
$\Theta$.

%%\subsection{Generic Approach to Parameter Estimation}
%%\label{sec:DPMM}

The aforementioned framework combines
$J$ distinct Gaussian regression models, and each model is weighted by parameter
$\pi_j$. To use this framework to describe the relationship between data,
parameters $\{\pi_{1:J}, \beta_{1:J}, \sigma^2_{1:J}\}$ need to be estimated.
In parameter estimation, the number of Gaussian regression models needs to be
determined. However, this number is typically unknown in advance.

To address this problem, one typical approach~\cite{Fraley01011998, ncomponents}
is to assign many possible
values to $J$, fit multiple mixture linear regression models and choose the
model based on Bayesian information
criterion (BIC), Akaike information criterion (AIC),
Deviance information criterion (DIC), or Bayes factor.
However, such an approach is not computationally efficient and may cause
under-fitting or over-fitting problems.

In this work, we therefore use a Dirichlet Process (DP)~\cite{dp,dp1,dp2}
prior to relax the restriction of $J$ and perform parameter estimation.
The basic idea of this approach is to
relax $J$ to infinity (i.e., $J \rightarrow \infty$). This is done by using DP to define the parameter set $\Theta$ over an infinite dimensional space, i.e., we allow an infinite number of
parameters a priori, and posterior inference is done to select the number of parameters, then
draw the posterior of $\Theta$, $P(\Theta|D)$, using Gibbs
sampler when conjugate priors are used. Note that $D$ in $P(\Theta|D)$ denotes the observed data.
In the following, we describe the detail of this approach.

Let a distribution $G$ follows a DP with parameters $\alpha$ and $G_0$,
denoted by $G \sim DP(\alpha,
G_0)$. This means $G$ is a (random) distribution, and we can draw samples from $G$ itself.
Here, $G_0$ is a base distribution -- which has a parametric form -- and
acts as a prior distribution over components parameters
$\{\beta_{1:\infty},\sigma^2_{1:\infty}\}$.
The concentration parameter, $\alpha$, is a positive scalar that controls the
variance of the DP. As $\alpha$ increases, $G$ is more likely to be
close to $G_0$, i.e., $G \rightarrow G_0$. Thus, $\alpha$ represents the degree
of confidence in the base distribution, $G_0$.

We use a ``stick-breaking" approach~\cite{citeulike:1495491} to explicitly construct the DP.
It is given as follows:
\begin{align*}
&\pi_j =  u_{j}\prod_{k=1}^{j-1}(1-u_k), j = 1,2,...,\infty ,\\
&\{\beta_j, \sigma^2_j\} \sim G_0,\\
& G = \sum_{j=1}^\infty \pi_j \delta_{\{\beta_j, \sigma^2_j\}}.
\end{align*}
Then $G \sim DP(\alpha, G_0)$, where  $\delta_{(\beta_j, \sigma^2_j)}$ is a point mass at $(\beta_j,\sigma^2_j)$, and
$\{u\}_{j=1}^{\infty}$ is a set of independent
beta distributed random variables such that $u_j \sim \text{B}(1, \alpha)$.

With the modeling of $\pi_j$, $\beta_j$ and $\sigma^2_j$ ($j =
1,2,...$) over the infinite dimensional space, we can use Markov Chain Monte Carlo (MCMC)
to draw their posterior
$P(\Theta|D)$ and estimate all the parameters in parameter set $\Theta$. As our
modeling shares the principle of Bayesian inference in which the priors are chosen to be conjugate,
we particularly use Gibbs sampling~\cite{10.1109/TPAMI.1984.4767596,
doi:10.1080/01621459.1990.10476213} to draw samples of $\Theta$ from its posterior distribution.

%%%%%%%%%%%%%%%%%%%%%%%%%%%%%%%%%%%%%%%%%%%%%%%%%%%%%%%%%%%%%%%%%%%%%%%%%%%%%%
%%\subsection{Estimating Parameters}
\subsection{Modeling}
\label{sec:EstimatingParameters}

% \comment{
% The aforementioned parameter estimation is only a generic approach. Using it
% to develop a specific model, we need to customize it and make it both
% computationally effective and efficient. In the following, we discuss how to
% customize the aforementioned parameter estimation for our problem.
% }

With the knowledge of Gaussian mixture model described above, we now define our problem with the following notation.
Consider a data set with $n$ observed 
measures $y_i$, $i = 1,...,n$, where each $y_i$ is an univariate
value representing the variation in speed gathered through the OBD-2 port of 
an automotive system. Corresponding to each observed acceleration measure, we also have an
unforgeable acceleration
measure $x_i$, where each $x_i$ is a
$3$-dimensional vector $x_i = (x_{i1}, x_{i2}, x_{i3})'$, representing an
acceleration measure in $3$-dimensional Euclidean space.
As is discussed earlier, the speed measures through a car's OBD-2 port can be
manipulated. Therefore, 
our goal is to identify $y_i$ manipulated by adversaries using $x_i$.

Following the aforementioned Gaussian mixture model, we 
model $y_i$ through $x_i$, and 
then have
\begin{align}
y_i \sim \sum_{j=1}^{J \rightarrow \infty}\pi_jN(y_i|x_i'\beta_j, \sigma^2_j),
\end{align}
and
\begin{align}
P(\Theta|D) \propto P(\Theta) \cdot \prod_{i=1}^{n} {\sum_{j=1}^{J \rightarrow \infty}\pi_jN(y_i|x_i'\beta_j, \sigma^2_j)},
\end{align}
with prior $P(\Theta)$ hierarchically defined as follows:
\begin{align}\label{eq:DPMGRM}
&\pi_1 = u_1, \text{ } \pi_j = (1-u_1)\cdots(1-u_{j-1})u_j, \text{ } 1<j < J,\\
&u_j|\alpha \sim \text{B}(1,\alpha), \text{ } j = 1,\dots, J-1,\\
& \alpha\sim G(e,f), \\
& \beta_j| \sigma^2_j \sim N(\mu_\beta, \sigma^2_j V_\beta),\\
&\sigma^2_j \sim IG(a,b),
\end{align}
for pre-specified hyperparameters $(e, f, \mu_\beta, V_\beta, a, b)$.
Here, $G(e,f)$ and $IG(a,b)$ represent the density of a Gamma distribution and
the density of an Inverse-Gamma distribution, respectively.

In our problem, we note that posterior
$P(\Theta|D)$ and prior $P(\Theta)$ are conjugate distributions. Thus, we can
draw $P(\Theta|D)$ using a Gibbs sampler. In particular, we use a standard
strategy in Bayesian computation. It augments the parameter space of the
statistical model through mixture component indicators $z_{1:n}$, i.e., $z_i = j$ indicates that $y_i$ was generated from mixture
component $j$, or $y_i | z_i = j, x_i  \sim N(y_i|x'_i\beta_j, \sigma^2_j)$, and
with $p(z_i = j) = \pi_j$. Hence, we can draw posterior and mixture component indicators jointly.

Considering drawing posterior $P(\Theta|D)$ is computationally inefficient,
especially when $J$ approaches infinity, we finally truncate $J$ by setting its
value to its upper bound. The truncation exploits the fact that
the number of components in a mixture regression model
cannot be greater than the number of observed data (i.e., $J \leq n$). The
reason behind this fact is that the construction
of mixture regression models follows a
process in which each observed data $(y_i, x_i)$ is assigned to one set of
regression coefficients with
the highest posterior probability. In other words, each pair of measure on acceleration
and speed variation is mapped to one regression component in a Gaussian mixture regression model. 

\subsection{Detecting Manipulation}
\label{subsec:detecting-manipulation}
Now, we discuss how to use the aforementioned Gaussian mixture model to identify fraudulent vehicle speed measures.

As is discussed in Section~\ref{sec:2attack}, we propose two manipulation strategies -- replaying
driving activities of a safe vehicle operator and altering dangerous driving
behaviors on
the fly. In fact, adversaries could come up with numerous
manipulation strategies not foreseen previously. It is not practical to use known manipulated data to
construct mixture model -- and use it to identify potential manipulation --
because we cannot obtain an encompassing data set with all the data manipulated
through all possible strategies.

As a result, we perform detection as follow. The mixture model takes an acceleration
measure  as an input and outputs a \preint. By examining if the speed
variation measured through a car's OBD-2 port lies in the range, we can
determine if the speed measure is manipulated or not.

In other words, given acceleration measure $x_i^*$ and variation in speed $y^*$
derived from two consecutive speed measures $y_i$ and $y_{i-1}$ acquired through a car's OBD-2 port, if
$y^*$ falls in the \preint $(a_i, b_i)$ derived from $x_i^*$, 
we identify speed measure $y_i$ normal
and unmanipulated. Otherwise, we identify it manipulated.
More specifically, we compute interval $(a_i, b_i)$ using the following
formula:
\begin{align}
p(y|x^*_i, D) &= \int p(y|x^*_i, \Theta)p(\Theta|D)d\Theta  \label{eq:pre}\\
%&\approx \frac{1}{S}\sum_{s=1}^Sp(y|x^*_i, \Theta_s) \\
& \approx \frac{1}{S}\sum_{s=1}^S \{\sum_{j=1}^J \pi_{j,s}N(y| {x^*_i}'\beta_{j,s}, \sigma^2_{j,s})\}.  \label{eq:preapp}
\end{align}
Where $D$ represents the data set that is used for model fitting.
Equation~\ref{eq:pre} is called the Bayesian predictive distribution.
Equation~\ref{eq:preapp} is a sampling based method for approximating the
predictive distribution, where $\{\pi_{1:J,s}, \beta_{1:J,s},
\sigma^2_{1:J,s}\}_{s=1}^{S}$ is a large set of samples obtained from the Gibbs
sampler. By generating a large set of samples from Equation~\ref{eq:preapp}, we
take a credible interval of the sample distribution as our \preint $(a_i, b_i)$.

\section{Evaluation}
\label{sec:eval}

In this section, we test our detection system in the real world. In particular,
we evaluate the effectiveness of our detection algorithm using a data set
collected from real drivers.

\subsection{Experimental Design}

To evaluate the effectiveness of our detection system, we design our experiment
as follows. We recruited volunteer drivers and separated them into two different
groups. In one group, we installed our detection system on cars and collected information about driving habits. 
Note that in our experiments, the data is collected from different car models.
The only requirement is that we held our system in the same
orientation to gather accelerations in a consistent setting\footnote{
In a real deployment, the insurance companies need to keep distinct 
classifiers for each car model because different car models have 
OBD-2 connectors in different orientations.}.
In the other group,
we plugged into cars our detection system along with the aforementioned MITM
box. As is discussed in Section~\ref{sec:attack}, the MITM box tampers abrupt
decrease in vehicle speed. Thus, the information acquired from the second group
contains manipulated measures on vehicle speed. 
We asked the second group of volunteers to intentionally apply hard-brakes now and then
to trigger the manipulation of OBD-2 speed readings\footnote{Volunteers only
hit the brakes in suburban area where there are less cars on the road.}.

\subsection{Experiment Configuration}

We conducted our experiment in the real world and configured our system as
follows.

\subsubsection{Hardware Configuration}

To collect driving information, we first configure our system prototype. Motor
vehicles from different manufacturers update their ECU modules at different
frequencies. While in gathering driving information, if our system requests ECU
modules slowly, we may not be able to capture the subtle change of vehicle
operation (i.e., lose sensitivity). In contrast, if our system requests driving
information at a high frequency, it will increase the redundancy of driving
information. In our experiment, we balance sensitivity and redundancy by
empirically setting our system to request driving information once every second.

Different from ECU modules built in cars, the accelerometer of our system can be
operated at a high frequency and detect the subtle change of vehicle operation.
In our experiment, we set the accelerometer with a sample rate of 16.7 Hz.

\subsubsection{Algorithm Configuration}

We also configure the parameters needed for our detection algorithm. As is
mentioned in Section~\ref{sec:model}, we generate posterior from prior
distributions. Here, we configure the parameters used by the prior
distributions. As we need the prior distributions to be less informative, we set
their parameters with $e = 10$, $f = 1$, $m = <0,0,0>$, $\lambda = 5$. We
initialize Gibbs sampler using our prior distributions and run for a total of
30,000 iterations. Then, we use the last $15,000$ draws as the posterior
distributions.

\subsection{Dataset}

\label{sec:dataset}
In total, our detection system collects 89 trips that constitute a set of raw
data covering 1,034 miles across vehicles from 6 manufacturers. Among the 89 trips,
40 trips contain manipulated vehicle speed. The manipulation to the OBD-2 speed reading is not triggered
frequently -- less than 1\% of all the collected data points are modified.
Each record in the raw data set
contains vehicle speed measured from a car's onboard diagnostic computer and
accelerations measured from accelerometers. An acceleration measure is a tuple
that contains three elements indicating the linear acceleration in three
orthogonal planes.

As is discussed earlier, our detection algorithm takes the input of acceleration
measures, outputs a \preint and examines if the variation in
consecutive speed measures belongs to the interval. Therefore, we compute the
speed variation by taking the difference of two consecutive measures on vehicle
speed. In addition, we re-construct the acceleration measures. As is mentioned
above, our system collects acceleration information at a high frequency while
gathering speed information at a relatively low rate. To utilize acceleration
measures to predict a speed variation, we need one-to-one mapping from
acceleration to vehicle speed. In our experiment, we take the average of all its
measures in a time window of one second, and then use the average measure to
represent the acceleration at that time point. In this way, we obtain a data set
with 90,434 records, each of which indicates a speed variation and the
corresponding accelerations at that time point.

As is mentioned above, we have two groups of volunteers who participate in our
experiment in different settings. Based on the source of the data acquired, we
therefore split our data records into two sets. In particular, our first set
contains 14,170 data records derived from the first group of volunteer drivers,
while the second comes from the other group of volunteers. As the first data set
does not contain tampered vehicle speed, we use it to construct our detection
model. In contrast, the second set contains manipulated vehicle speed, and thus
we use it to evaluate our detection model.

\subsection{Results}

Using the algorithm described in Section~\ref{sec:model} and the data set
discussed above, we construct a mixture Gaussian regression model. In the
following, we describe the characteristics of the model followed by the
evaluation of its effectiveness.

\subsubsection{Characteristics}

\begin{figure*}[t!]
    %\vspace{-6mm}
    \centering
    \begin{subfigure}[t]{0.3\textwidth}
        \centering
        \includegraphics[width=1\columnwidth]{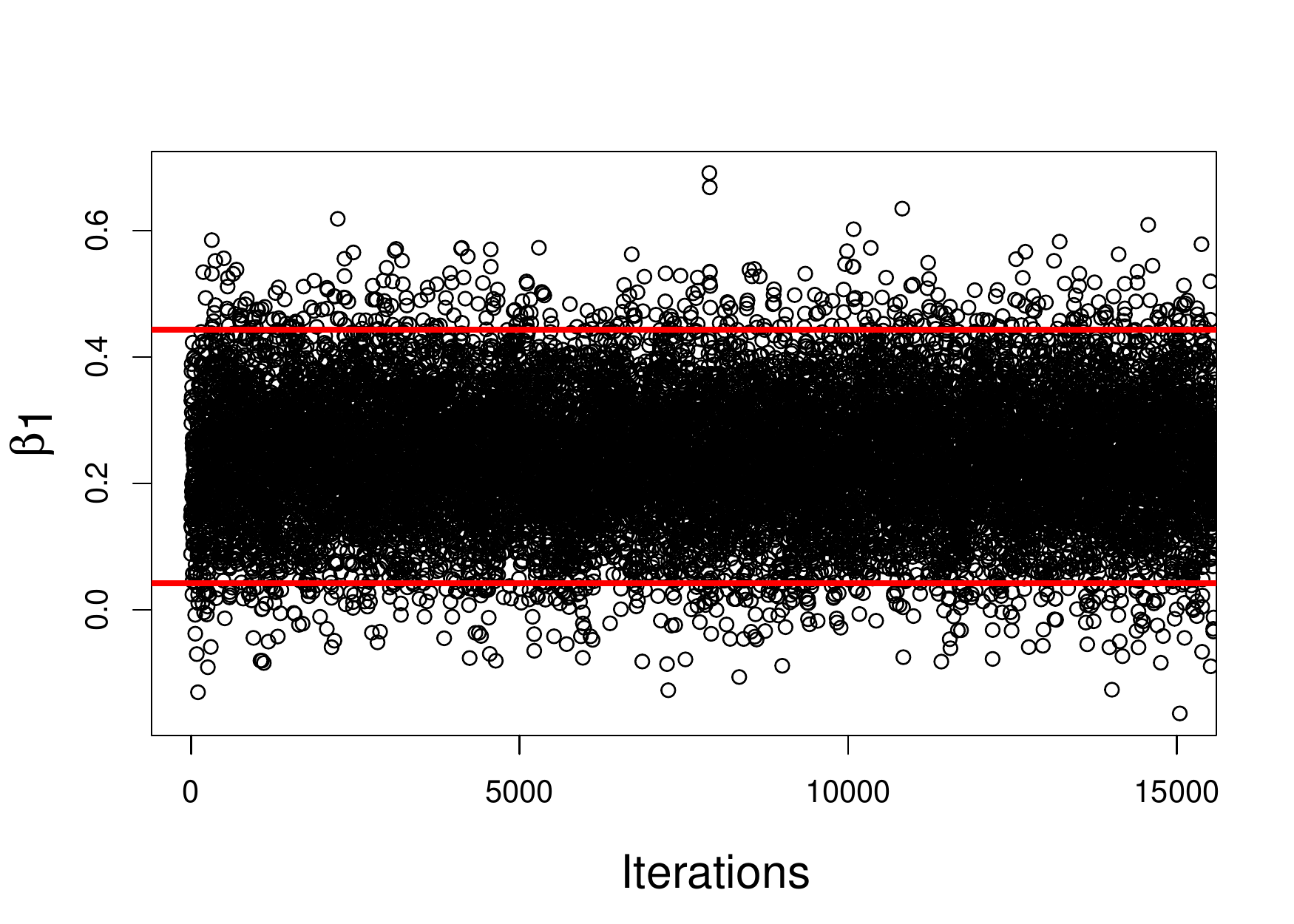}
        \caption{Forward}
        \label{fig:contributionbeta1}
    \end{subfigure}%
    ~
          \begin{subfigure}[t]{0.3\textwidth}
        \centering
        \includegraphics[width=1\columnwidth]{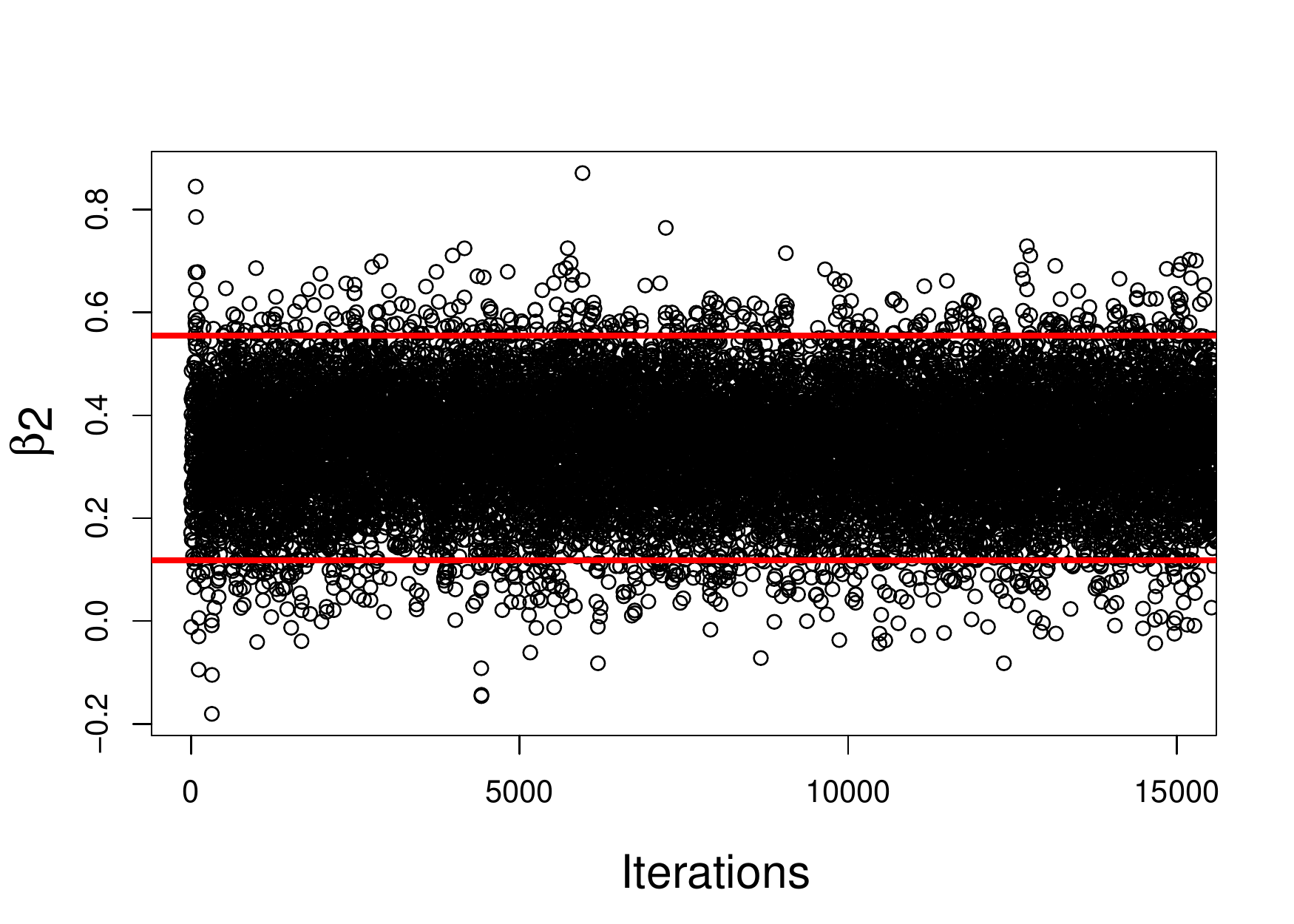}
        \caption{Lateral}
        \label{fig:contributionbeta2}
    \end{subfigure}
    ~
    \begin{subfigure}[t]{0.3\textwidth}
        \centering
        \includegraphics[width=1\columnwidth]{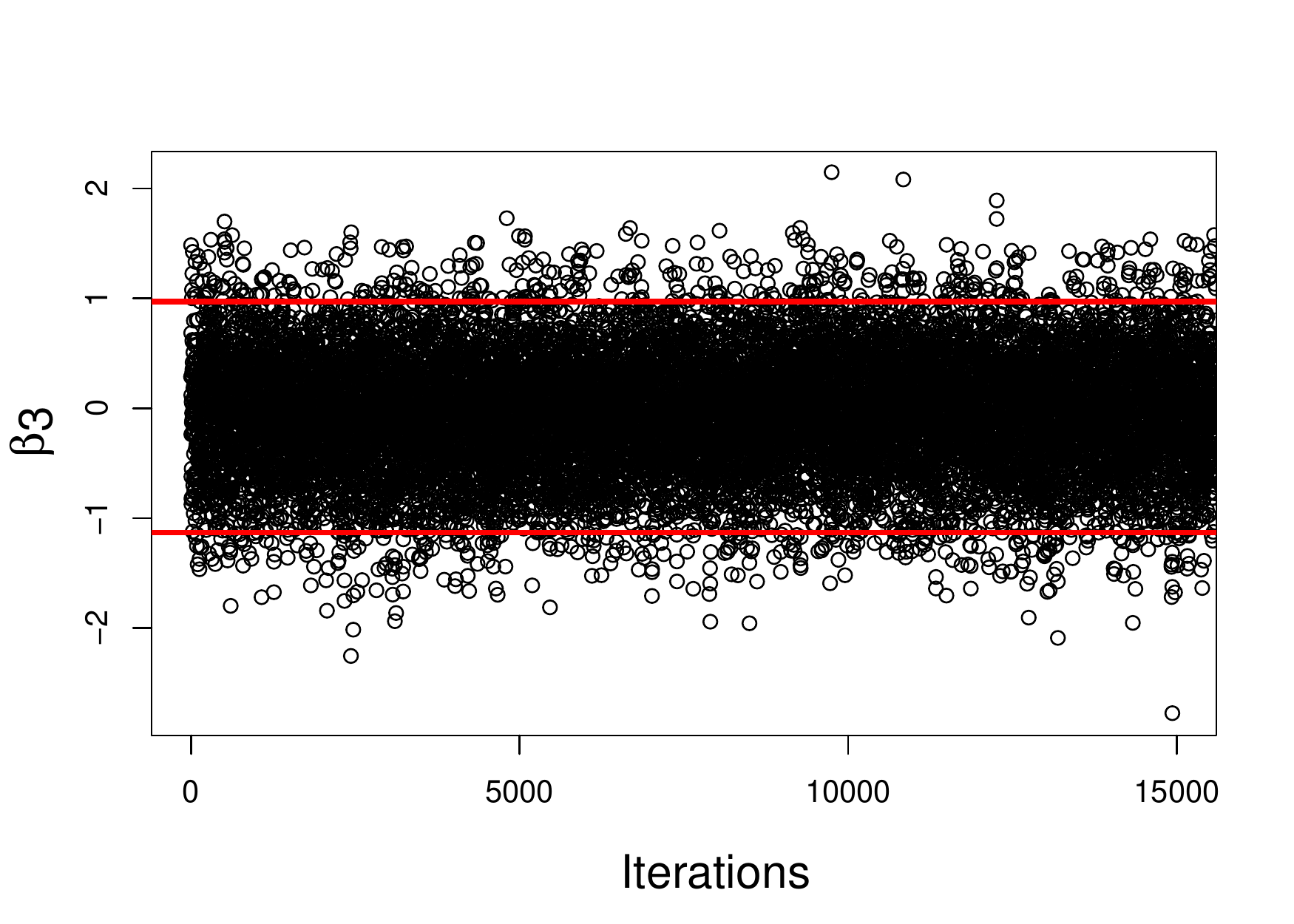}
        \caption{Perpendicular}
        \label{fig:contributionbeta3}
    \end{subfigure}
   \caption{\label{fig:contribution}The influence of the acceleration in each direction.}
   % \vspace{-4mm}
\end{figure*}

\begin{table}
\centering

\begin{threeparttable}
\small{
\caption{10 most significant components contributing the mixture regression model}
\label{tab:pi}
\begin{tabular} {c||ccccc}
\hline
i&1&2&3&4&5\\
$\pi_{i}$ &0.5020&0.0788&0.0780&0.0742&0.0568 \\ \hline  \hline
i&6&7&8&9&10 \\
$\pi_{i}$ & 0.0423&0.0396&0.0370&0.0360&0.0331 \\
\hline
\end{tabular}
}
\end{threeparttable}
\vspace{-4mm}
\end{table}

Our final mixture regression model contains 10 ``non-empty" components, i.e.,
$\pi_i \geq 10^{-5}$ with $i \in [1,2,\cdot\cdot\cdot 10]$. Table~\ref{tab:pi}
shows the value of the weight across each component (i.e., $\pi_{1:10}$). As we
observe from the table, our model consists of one dominant and nine minor
components because the value of $\pi_1$ is significantly greater than the values
of the succeeding weights. This indicates that, the variation in vehicle speed
is mainly dominated by the accelerations of a motor vehicle but at the same time
largely influenced by unmeasurable noise (e.g., the noise introduced by poor
road condition, engine vibrations and bad weather).

Taking a close look at each component -- especially the values of parameter
$\beta_{1:3}$ -- we further examine the most and least significant variables in
our model. To illustrate, we take the dominant component for example and plot
the draws of $\beta_{1:3}$ in Figure~\ref{fig:contribution}. In this figure, the
space between the red lines represents a 95\% credible interval. Using it, we
can easily assess both the practical and statistical significance for a given
covariate. In this particular case, we can say variable $x_1$ and $x_2$ are
significant because the central points of $\beta_1$ and $\beta_2$ fall at 0.2
and 0.4, respectively. In addition, both the credible intervals do not contain
$0$. In contrast, $x_3$ is a least significant variable for the reason that the
central point of $\beta_3$ is approximately zero. In our experiment, $x_1$ and
$x_2$ represent the accelerations in forward and lateral directions, whereas
$x_3$ indicates the acceleration in the direction perpendicular to the ground.
As such, variable $x_3$ has nearly no contribution to speed variation, whereas
variable $x_1$ and $x_2$ are more informative in predicting speed variation.

\subsubsection{Effectiveness}

\begin{figure}[t!]
    \centering
    \begin{subfigure}[t]{0.42\textwidth}
        \centering
        %\vspace{-8mm}
        \includegraphics[width=1\columnwidth]{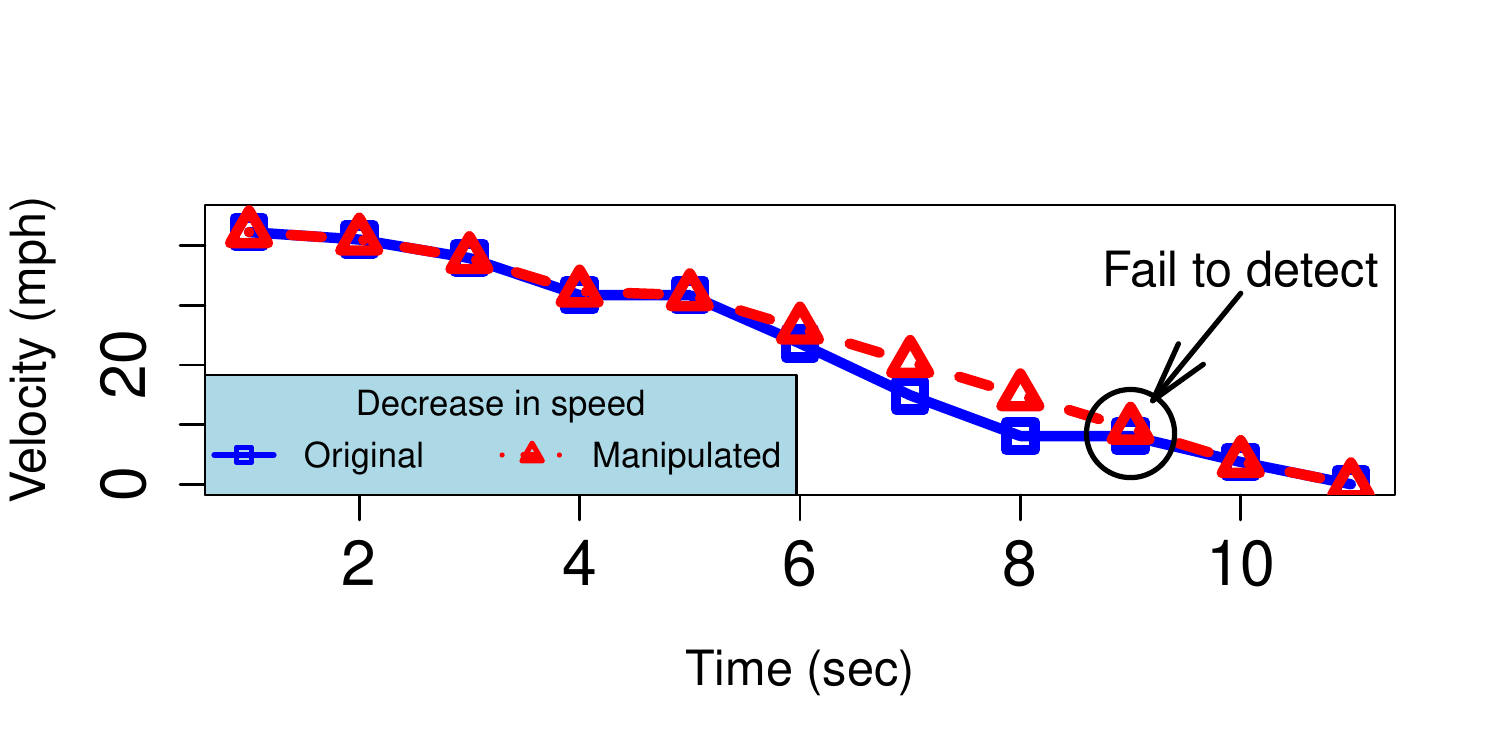}
        %\vspace{-8mm}
        \caption{False negative.}
        \label{fig:fnanalysis}
    \end{subfigure} %\\
    ~
    \begin{subfigure}[t]{0.42\textwidth}
        \centering
        %\vspace{-8mm}
        \includegraphics[width=1\columnwidth]{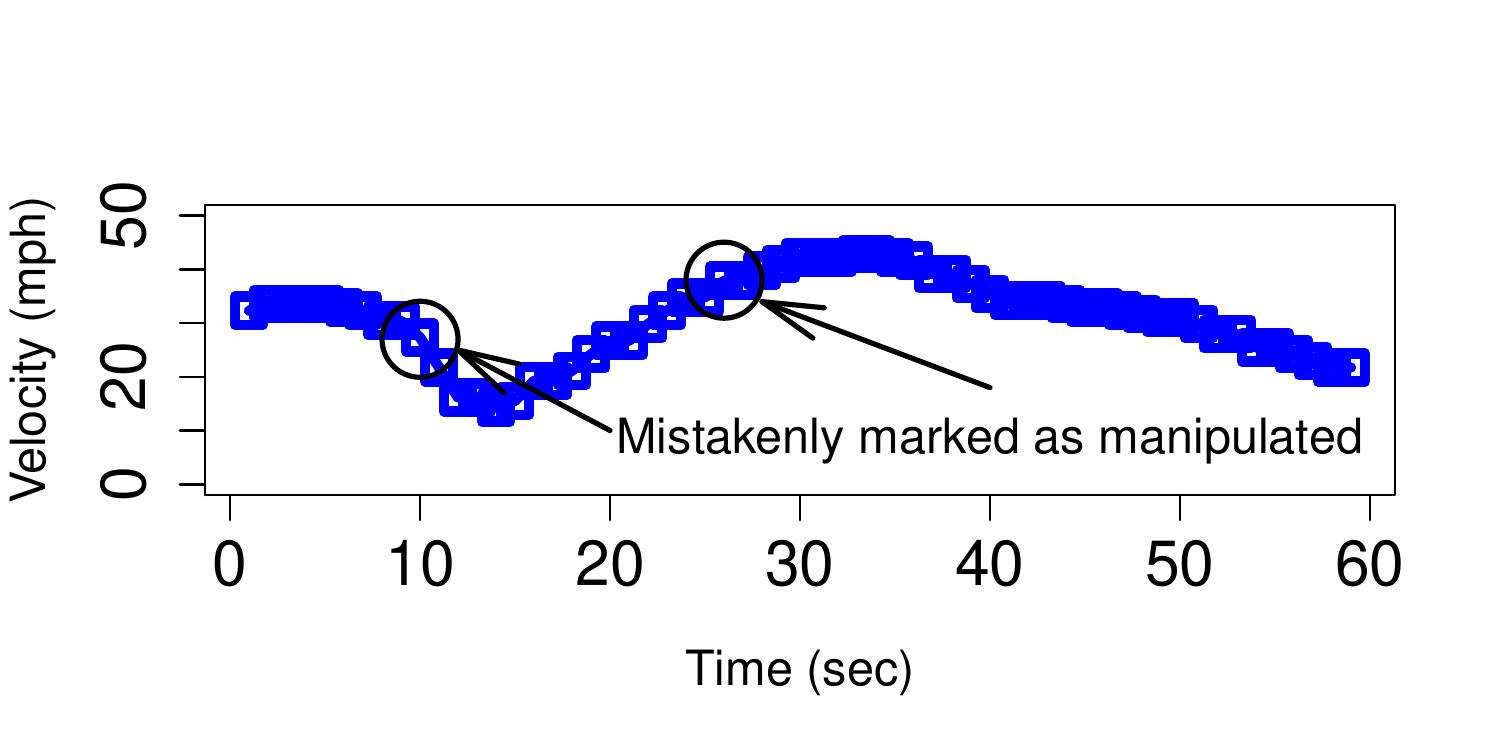}
        %\vspace{-8mm}
        \caption{False positive.}
        \label{fig:fpanalysis}
    \end{subfigure}
    \caption{Analysis of detection failures.}
    %\vspace{-3mm}
    \label{fig:fnfpanalysis}
\end{figure}

\begin{figure}
\centering
  \includegraphics[width=0.8\columnwidth]{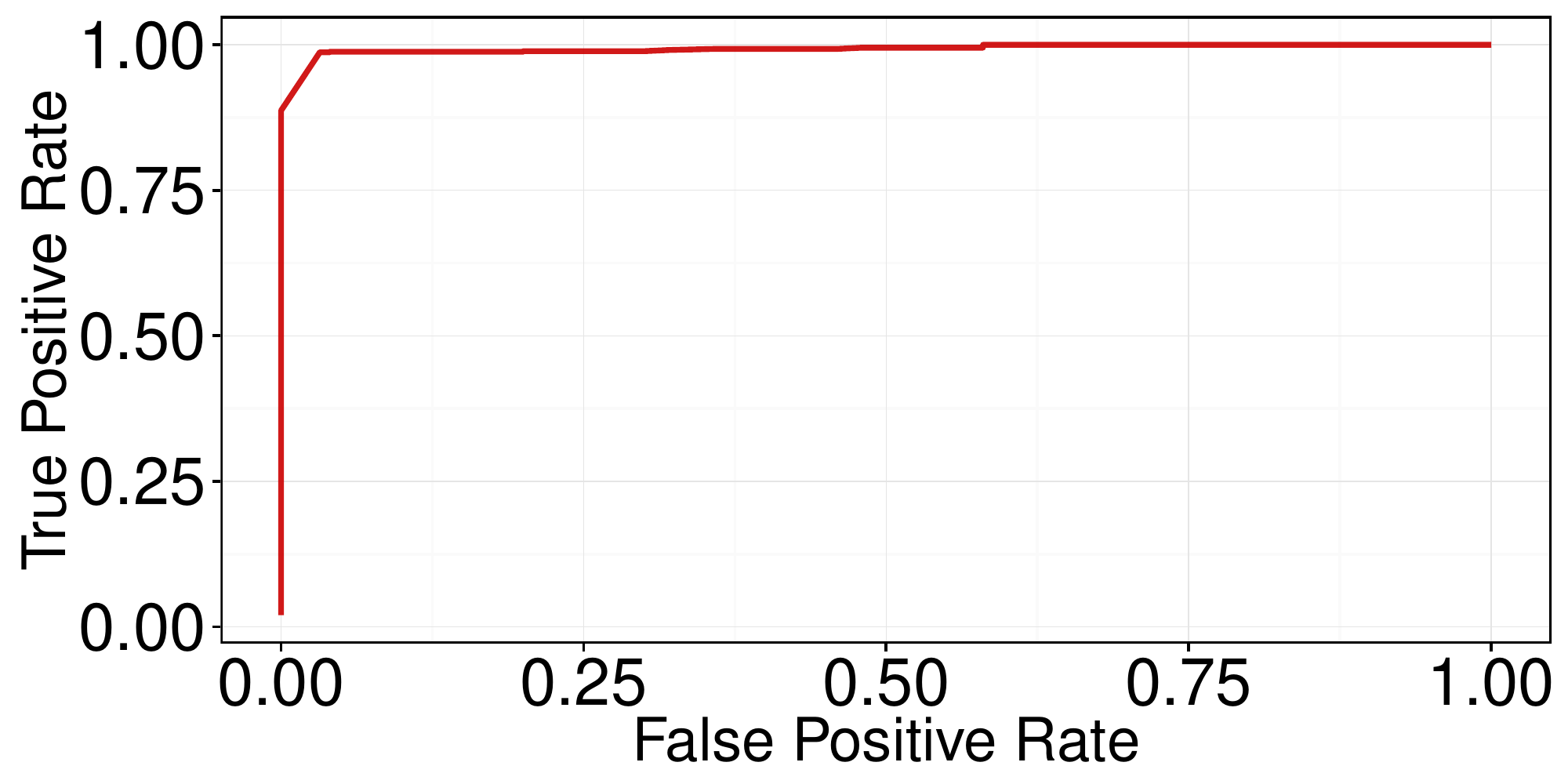}
  \caption{The ROC curve of our detection mechanism.}
  \label{fig:roc}
  %\vspace{-6mm}
\end{figure}

Figure~\ref{fig:roc} plots a receiver operating characteristic (ROC) curve
that illustrates the true positive rate against the false positive rate at 
various threshold settings.
Our detection algorithm exhibits a false positive rate of 0.032 when the false
negative rate is equal to 0.013. This low false negative indicates that our
detection is highly precise in identifying manipulated vehicle speed, and the
low false positive demonstrates that our detection algorithm is less likely to
pinpoint an unmanipulated speed as a manipulated one.

We also take a closer look at those small amounts of manipulated data points that
our algorithm fails to detect. Figure~\ref{fig:fnanalysis} showcases a
representative sample data trace indicating a hard brake before and after it
gets manipulated. Our detection algorithm can identify all the manipulated data
points except the one marked in the figure.  Although our algorithm fails to
identify the last manipulated data point, we observe that, the algorithm accurately
marks the majority of manipulated data points in this dangerous hard brake. 
If we take this hard brake as an event, 
this false negative is by no means our
algorithm misses out speed manipulation.
As such a case is common in our data set, our algorithm can successfully identify all
hard brake events despite missed individual data points.

Finally, we inspect those data points that our algorithm mistakenly identifies
as manipulated. In Figure~\ref{fig:fpanalysis}, we illustrate a representative
unmanipulated data trace in which we highlight two data points that we mistakenly
identify as manipulated. Compared with those manipulated data points identified
precisely in Figure~\ref{fig:fnanalysis}, the mistakenly identified data points
are spread out across time. This means we can easily rule out the data points
that our algorithm mistakenly identifies as manipulated (i.e., false positive).
Similar to the showcase discussed above, such an observation is common in our
data set, and thus the false positives do not incur misjudgment (i.e.,
identifying an innocent driver as a miscreant).

\section{Discussion}
\label{sec:discussion}

In this section, we first discuss some security and overhead issues of the proposed
defense mechanism.  
Then, we discuss other defense options, their limitations and some
related issues concerning the aforementioned attack.

\para{Issues Related to the Current Design.}
Our statistical model outputs a \preint. 
In detecting speed manipulation, we examine if the speed variation measure
from the OBD-2 extends outside our \preint.
A sophisticated attacker could implement an enhanced MITM box that
manipulates the OBD-2 readings based on the \preint.
However, such attack is infeasible considering the restricted computing capability
in the telematics devices.

In our design, the telematics devices additionally send averaged accelerometer readings to
the insurer's server, which may have impacts to both the computation and network communication.
Since the averaging algorithm mentioned in Section~\ref{sec:dataset} is extremely simple, the
computation overhead can be neglectable.
In addition, the averaged acceleration data is sent at 1HZ along with the OBD-2 readings,
so the network traffic is quadrupled (acceleration data has 3 axes).

\para{Spatial Anomaly Detection.}  In addition to the aforementioned defense
mechanism, another defense option is to detect manipulation using location
information, which can be obtain from GPS chips~\cite{Lu:2003:ASO:951949.952103}.
% As is mentioned earlier, some insurance telematics devices have
% already integrated GPS components. This gives an insurer the ability to use
% spatial anomaly detection mechanisms~\cite{Lu:2003:ASO:951949.952103} to
% identify unlawful velocity manipulation. More specifically, 
An insurer can use
GPS information to identify the motion of the car at a coarse granularity, and
catch fraudulent activities if a miscreant simply replays a  pre-recorded
driving trace without making his vehicle in motion.

However, the location based defense provides inadequate reliability and
accuracy. From a reliability standpoint, the location information may not be
always available. Since the GPS receiver is dependent upon unobstructed view to
satellites, the location based detection fails when a vehicle is moving along
urban canyons (tall buildings and tunnels), 
or a GPS device is unable to stay sufficiently close to the
window of a moving vehicle. From an accuracy standpoint, current GPS-based
solutions provide only modest accuracy. 
GPS devices are positional speedometers based on how far the device has moved
since the last measurement. 
The speed information is calculated based on formulas and 
is normalized, so it is not an instant value~\cite{serrano2004single}. 
Many GPS receivers output data at 1Hz, as a result, 
fine-grain acceleration data in this second is lost. 
This coarse-grain measurement is ineffective for subtle manipulates of
velocity readings.
On the contrary, using
accelerometer, we can capture instant acceleration at a higher frequency (in
our experiment, we used 16.7 HZ).

\para{Driving Behavior Identification.}  Abnormal driving behavior monitoring is
a technique  developed to improve drivers' awareness of their driving habits
(e.g.,~\cite{chensecon15, Paefgen:2012:DBA:2406367.2406412}). Using a variety of
sensors, such as accelerometer and gyroscope, it identifies dangerous driving
activities in a trip. In customizing insurance rate, insurers also identify
dangerous driving activities, particularly using the velocity readings acquired
from a telematics device. As such, intuition might suggest that the driving
behavior monitoring technique could be a substitution to current 
telematics-based technique. Since existing tamper-resilient techniques can safeguard
sensors and make sensor readings unforgeable~\cite{sensorintegrity}, this substituted solution can be 
 another defense option. 
For example, AAA Drive is an APP that is installed on a customer's smartphone to 
monitor the driver's driving behavior~\cite{aaadrive}.
Similar to telematics-based techniques, good driving behavior is the criterion for
premium discount.
In practice, this solution however is infeasible. 
First, smartphone-based monitoring cannot stop a miscreant to refuse to log high-risk
journeys.
Currently, insurer like AAA requires the customers to maintain a minimal mileage each
month to remain eligible for the program.
Second, insurers  need to identify subtle variations in vehicle speed to determine
if a vehicle brakes suddenly, whereas  the driving behavior monitoring cannot
accurately distinguish information pertaining to movement behavior from other
factors that affect the sensor readings. In particular, potholes and other
severe road surface can mask the relevant information.

% \para{Attacker's effort and risk}  As is discussed earlier, our attack uses
% a small piece of program and off-the-shelf hardware to fabricate and manipulate
% velocity readings that an insurance telematics device collects. As such, this
% requires little efforts for a miscreant to obtain unlawful financial profits.

% Since the telematics device is at the end users' disposal, which imposes
% additional difficulties on the insurers in monitoring when and how their device 
% is used, a miscreant can launch the aforementioned attack with minimal risk of
% being exposed. For example, the miscreant can use the aforementioned
% man-in-the-middle box to manipulate velocity readings subtly and, without our
% defense mechanism, the insurer cannot determine the legitimacy of the velocity
% readings.

\section{Conclusion}
\label{sec:conclusion}

In this paper, we present a new attack against personalized auto insurance services.
It exploits the fact that personalized auto insurance services are lack of an effective
mechanism to verify the authenticity of the data that they collect for
personalized pricing. Launching this attack, a miscreant can obtain unlawful
financial profits.
While the attack is trivial, it can bring about significant negative
impacts on auto insurers, especially considering an increasing number of
auto insurers use personalized pricing to open up new avenues of growth.
In response to the new attack, we also introduce a defense
mechanism which verifies data integrity by utilizing unforgeable data acquired
from the physical world.
Using trustworthy data acquired
from the physical world, we construct a sophisticated statistical model to
counteract the attack against personalized insurance pricing. As part of future
work, we plan to collaborate with auto insurers and incorporate our defense
mechanism into their personalized auto insurance services.

\section{Acknowledgments}
We would like to thank the anonymous reviewers for their helpful feedback and
our shepherd, Marco Zuniga, for his valuable comments on revision
of this paper. This work is partially supported by U.S. Army Research Office
under Grant No. W911NF-13-1-0421 (MURI), by National Science
Foundation under Grants No. CNS-1505664 and by Penn State Institute
for Cyber Science Seed Funding Initiative grant. Any opinions,
findings, and conclusions or recommendations expressed in this
material are those of the authors and do not necessarily reflect the
views of the National Science Foundation and U.S. Army Research Office.

%\small
\balance{
\bibliographystyle{abbrv}
\bibliography{priv,sslab,conf,snapshot,jx,ref,sensys}} % sigproc.bib is the name of the Bibliography in this case

\end{document}